\documentclass[superscriptaddress,aps,pra,twocolumn,showpacs,nofootinbib,longbibliography,showkeys]{revtex4-2}
\usepackage{amsmath,amssymb,amsthm}
\usepackage{easybmat}
\usepackage[colorlinks=true,citecolor=blue,urlcolor=blue, linkcolor = magenta]{hyperref}
\usepackage[pdftex]{graphicx}
\usepackage{times,txfonts}
\usepackage{braket}
\usepackage{color}
\usepackage{natbib}
\newcommand{\be}{\begin{equation}}
	\newcommand{\ee}{\end{equation}}
\newcommand{\ba}{\begin{eqnarray}}
	\newcommand{\ea}{\end{eqnarray}}
\newcommand{\ketbra}[2]{|#1\rangle \langle #2|}

\begin{document}
	
	%\preprint{APS/123-QED}
	\title{Impact of non-Markovian evolution on characterizations of quantum thermodynamics}

	\author{Devvrat Tiwari\textsuperscript{}}
	\email{devvrat.1@iitj.ac.in}%Lines break automatically or can be forced with \\
	\author{Subhashish Banerjee\textsuperscript{}}
	\email{subhashish@iitj.ac.in }
	\affiliation{Indian Institute of Technology Jodhpur-342030, India\textsuperscript{}}
	
	%\textsuperscript{b,c}}
%\affiliation{
	%Third institution, the second for Charlie Author
%\affiliation{%


\date{\today}% It is always \today, today,

\begin{abstract}
	Here we study the impact of non-Markovian evolution on prominent characteristics of quantum thermodynamics, such as ergotropy and power. These are benchmarked by the behavior of the quantum speed limit time. We make use of both geometric-based, particularly quantum Fisher and Wigner-Yanase information metric, and physical properties based-measures, particularly relative purity measure and relative entropy of coherence measure, to compute the quantum speed limit time. A simple non-Markovian model of a qubit in a bosonic bath exhibiting non-Markovian amplitude damping evolution is considered, which, from the quantum thermodynamic perspective with finite initial ergotropy, can be envisaged as a quantum battery. To this end, we explore the connections between the physical properties-based measures of quantum speed limit time and the coherent component of ergotropy. The non-Markovian evolution is shown to impact the recharging process of the quantum battery. Further, a connection between the discharging-charging cycle of the quantum battery and the geometric measures of quantum speed limit time is observed. 
\end{abstract}

%Use showkeys class option if keyword
\maketitle

%\tableofcontents
\section{Introduction}\label{sec-intro}
A realistic quantum system is subjected to the influence of the environment. The dynamics of the system is altered by this, and as a consequence, information is lost from the system to the external environment. A paradigm for examining how the ambient environment affects a quantum system is provided by the theory of open quantum systems (OQS)~\cite{bruer-petrrucione, sbbook, weiss}. Ideas pertaining to open quantum systems are applicable to a number of scenarios~\cite{Louisell, caldeira-leggett1983, GrabertSchrammIngold, sbqbm, sbsterngerlach, sbrichard, sbjavidprd, sbkhushboocoherence, sbunruh1, sbunruh2, reactioncoordinaterefs1, reactioncoordinaterefs2, reactioncoordinaterefs3, blhu, sgad, plenio}. 
A Markovian approximation, which implies that the environment instantly recovers from its contact with the system, can be used to describe the evolution of an OQS in many situations. This results in a constant movement of information from the system to the environment. However, research has been pushed into domains beyond the Markovian evolution due to increasing technical and technological advancements. In many of these situations, a clean division between system and environment time scales cannot be anticipated, leading to non-Markovian behavior~\cite{breuer1, Breuer_2012, rivas, rivas1, vasile,lu,luo, fanchini, haseli,hall1,sss_dephasing_depolarizing, sam1-banerjee, devega-alonso, laine, kumar2018non, nmad-channel, phase-covariant, Hakoshima, xingli_li}. Non-Markovian behavior, such as that caused by strong system-bath coupling, can delay decay and sometimes even cause a rebirth of quantum effects~\cite{kumar2018non, Wang_2013, devvrat_qsl}. The dynamics of quantum speed limit time, introduced below, can demonstrate how the evolution of the system of interest might alter owing to the nature of the bath~\cite{Deffner2_2017, Pfeifer}.

One of the cornerstones of quantum physics is Heisenberg's uncertainty principle. The uncertainty principle for position and momentum demonstrates that it is impossible to measure both position and momentum at the same time accurately; however, the meaning of the energy-time uncertainty relation in this statement is not immediately clear. The energy-time uncertainty principle ($\Delta t \ge \hbar /\Delta E$) is a statement about the intrinsic time in which a quantum system develops, as demonstrated by Mandelstam and Tamm (MT) in~\cite{mandelstam1945}. Using the Fubini-Study metric on the space of quantum pure states, the concept of the quantum speed limit (QSL) time, or the speed at which a quantum system can evolve, was first presented in~\cite{anandan_aharanov}. Margolus and Levitin (ML)~\cite{MARGOLUS} gave a different QSL time depending on the mean energy. Combining the MT and ML bounds results in a tighter QSL time for unitary dynamics, limited to orthogonal pure states. QSL time alludes to the shortest period of time needed for a quantum system to evolve from one state to another. It plays a crucial role in the development of quantum technologies such as quantum computing and quantum communication. Over the years, many measures to quantify the QSL time have emerged. Here we use the geometric-based measures, particularly Fisher and Wigner-Yanase (WY) information metrics-based measures, as well as the inherent dynamics-based measures, particularly relative entropy, and entropy of coherence-based measures, to compute the QSL time.
%Any actual system that interacts with its surroundings risks losing its quantum coherence. The concepts of OQS can be used to account for this. 
The design and execution of quantum information processing algorithms are significantly impacted by the expansion of the QSL time to open quantum systems~\cite{deffner2013qsl, qsl_relative_purity, Deffner2_2017, adesso}. Due to its applicability to other technical areas, this has been a subject of considerable recent research~\cite{action_qsl, riya_qsl_phase_covariant, devvrat_qsl, Paulson_role_coherence, Aggarwal_2022, Wei2016, qsl_pati1, haseli_2}. Further, the idea of the lower bound for the time required to transform an initial state to a final state in a non-Markovian environment is an important area to explore from the perspective of quantum thermodynamics~\cite{Arpan_Das_2021, Funo_2019}.

The fundamental laws of thermodynamic equilibrium and nonequilibrium in the quantum regime are the subject of quantum thermodynamics~\cite{gemmer2004quantum, binder2019thermodynamics, deffner2019thermo, sai_quant_thermo, Ali2020}. The emerging field of quantum thermodynamics has grown rapidly over the last decade. The main objective of quantum thermodynamics is to extend classical thermodynamics to incorporate quantum effects and tiny ensemble sizes. This is facilitated by the rapid experimental control of quantum systems and the engineering of small environments. The effect of memory on a quantum thermodynamic system has been a recent area of interest~\cite{thomas_heat_engine, whitney_2018, czartowski2023}. The extraction of maximal work from a system is an old problem in thermodynamics. It was shown~\cite{Allahverdyan_2004, cakmak_ergo} that in quantum systems, this can be quantified by the ergotropy of the system. The ergotropy has been established as an important quantity in the emerging field of quantum thermodynamics~\cite{kosloff_2013, Goold_2016, Mitchison, Francica_coherent_ergo, Deffner_coherent_ergo} and has recently been measured experimentally~\cite{VanHorne2020, von_lindenfels}. 

Quantum batteries are quantum-mechanical energy storage devices~\cite{alicki, binder2019thermodynamics}. The role of quantum effects on the issue of energy storage has been extensively studied in the last years~\cite{ferrar0_2018, Binder_2015, Levinsen_2018, Andolina_2018, battery_optimal, aditi_sen1, aditi_sen2}. The problem of quantum battery's charging and discharging has been studied in an open quantum system setting~\cite{Donato_2019} along with the impact of non-Markovian evolution~\cite{thomas_heat_engine, Kamin_2020}. In recent years, research on quantum batteries has gained much attention, and various methods have been employed to realize a practical quantum battery~\cite{Seah_collision_battery, Salvia_2023}. To this end, we aim to study a simple OQS model exhibiting non-Markovian behavior, {\it viz.}, a non-Markovian amplitude damping (NMAD) evolution. We exploit characterizers of quantum thermodynamics, particularly ergotropy, along with instantaneous and average powers, to study this model from the perspective of a quantum battery. This is benchmarked by the QSL time of the evolution of the system, where we use geometric and inherent dynamics-based measures of QSL time. 

This paper is organized as follows. In Sec. \ref{sec-Prelim}, we discuss the preliminaries used throughout the paper, including the discussion of QSL time, an OQS model with NMAD evolution, and ergotropy, instantaneous and average powers. Section \ref{sec-coherent-connect} discusses a direct connection between the physical properties-based measure of QSL time and the coherent component of the ergotropy. We study the connection of geometrical based-measures of QSL time with ergotropy, instantaneous and average powers, and impact of non-Markovian evolution over the discharging-charging process of quantum batteries in Sec. \ref{sec-qsl-ergo-connect}. This is followed by conclusions. 
\section{Preliminaries}\label{sec-Prelim}
This section briefly discusses different bounds on the quantum speed limit time, a single qubit model with non-Markovian evolution modeled by the non-Markovian amplitude damping (NMAD) master equation, and ergotropy followed by instantaneous and average powers.

\subsection{Quantum speed limit (QSL) time}
Here we discuss four forms of the quantum speed limit (QSL) time. The first two are related to the geometric QSL time, where we use quantum Fisher information and Wigner Yanase (WY) information metrics for the geodesic distance between the initial and the final state at time $t$. The other two measures depend on the inherent dynamics of the system, one of which is a definition of QSL time using the relative purity measure of a quantum state between initial and final states, and the other is a QSL time based on the coherence of the initial and final states. 

\subsubsection{Using quantum Fisher information metric}
Mandelstam and Tamm (MT)~ and Margolus and Levitin (ML)-type bounds on speed limit time~\cite{mandelstam1945, MARGOLUS} are estimated by using the geometric approach, using the Bures angle, to quantify the closeness between the  initial and final states. 
This approach was used to provide a bound for the initial pure state $\rho_0=\vert\psi_0\rangle\langle\psi_0\vert$, on the quantum speed limit time $(\tau_{QSL})$~\cite{deffner2013qsl} as 
\begin{equation}
    \tau_{QSL}=\max\Bigg\{\frac{1}{\Lambda^{\textrm{op}}_{t}},\frac{1}{\Lambda^{\textrm{tr}}_{t}},\frac{1}{\Lambda^{\textrm{hs}}_{t}}\Bigg\} \sin^2[\mathcal{B}],
    \label{eq-tau_qsl_Fisher}
\end{equation}
where  $\mathcal{B}(\rho_0,\rho_t)=\arccos\left(\left\{{\rm Tr}\left[\sqrt{\sqrt{\rho_0}\rho_t\sqrt{\rho_0}}\right]\right\}^2\right)$, and
\begin{equation}
    \Lambda^{\textrm{op,tr,hs}}_{t}=\frac{1}{t}\int^{t}_{0} ds \vert\vert \mathcal{L}(\rho_s)\vert\vert_{\textrm{op,tr,hs}}.
    \label{eq_qsl_Fisher_norms}
\end{equation}%  
The three norms $\vert\vert \cdot \vert\vert_{\text{op}}$, $\vert\vert \cdot \vert\vert_{\text{tr}}$, and $\vert\vert \cdot \vert\vert_{\text{hs}}$ are the operator, trace, and Hilbert-Schmidt norms, respectively. $\mathcal{L}$ is the Liouvillian superoperator acting on $\rho$. From the norm inequalities, it can be shown that the operator norm of the generator provides a tighter bound on the quantum speed limit time.

\subsubsection{Using Wigner-Yanase information metric}
Here we make use of the Wigner-Yanase information metric given by 
\begin{equation}
    \mathcal{B}(\rho_0, \rho_t) = \arccos\left({\rm Tr} \left[\sqrt{\rho_0}\sqrt{\rho_t}\right]\right),
    \label{eq-wy_metric}
\end{equation}
in the expression of the QSL time given in Eq. (\ref{eq-tau_qsl_Fisher}). The QSL time obtained using this metric is $\tilde \tau_{QSL}$. In~\cite{adesso}, it was shown that using this metric, one can get an upper bound on the QSL time in the case of mixed states. 

\subsubsection{QSL time using relative purity measure}
A bound on the QSL time for the open quantum systems can be given based on the relative purity measure given in~\cite{qsl_relative_purity}. The bound to the required time of evolution, in this case, is given as 

\begin{equation}
    t \ge \tau'_{QSL} = \frac{4\theta^2{\rm Tr} (\rho_0^2)}{\pi^2\overline{\sqrt{{\rm Tr}\left[\left(\mathcal{L}^\dagger \rho_0\right)^2\right]}}},
    \label{eq-qsl_rel_purity}
\end{equation}
where $\theta = \cos^{-1} [f(t)]$. $f(t) = {\rm Tr}(\rho_t\rho_0)/{\rm Tr}(\rho_0^2)$ is the relative purity measure, where $\rho_0$ is the initial state and $\rho_t$ is the state evolved to time $t$. In the denominator, we have $\Bar{(\cdot)} = t^{-1}\int_0^{t}(\cdot) ds$.

\subsubsection{QSL time for the coherence}
There are several widely known (basis dependent) quantum coherence measures, such as the relative entropy of coherence, the $l_1$ norm of coherence, the geometric coherence, and the robustness of coherence. The relative entropy of coherence is used because of its operational meaning as distillable coherence. In addition, it is also easier to work and compute compared to some other measures of coherence. For a given state $\rho$, the relative entropy of coherence $C(\rho)$ deﬁned as 
\begin{equation}
    C(\rho) = S(\rho^D) - S(\rho),
    \label{eq-coherence_1}
\end{equation}
where $\rho^D = \sum_i \bra{i}\rho\ket{i}\ketbra{i}{i}$ is the density operator that is diagonal in the reference basis, obtained by dephasing off-diagonal elements of $\rho$, and $S(\rho) = -{\rm Tr}[\rho\log\rho]$ is the von Neumann entropy of the state $\rho$.

For an arbitrary quantum dynamics of a ﬁnite-dimensional quantum system describable as the time-evolution of its state, the minimum time needed for the state $\rho_{t}$ to attain coherence $C(\rho_t)$, starting with the initial coherence $C(\rho_0)$, is lower bounded by quantum speed limit time for coherence ($\tau_{CSL}$) given by~\cite{qsl_pati1}
\begin{equation}
    t \ge \tau_{CSL} = \frac{\vert C(\rho_{t}) - C(\rho_0)\vert}{\Lambda^{\text{rms}, D}_{t} \overline{\vert\vert\ln \rho_s^D\vert\vert^2_{HS}} + \Lambda^{\text{rms}}_{t}\overline{\vert\vert\ln \rho_s\vert\vert^2_{HS}}},
    \label{eq-qsl_coherence}
\end{equation}
where $\Lambda^{\text{rms}, D}_{t} = \sqrt{\frac{1}{t}\int_0^{t} \vert\vert\mathcal{L}_s(\rho_s^D)\vert\vert^2_{HS} ds}$, $\Lambda^{\text{rms}}_{t} = \sqrt{\frac{1}{t}\int_0^{t} \vert\vert\mathcal{L}_s(\rho_s)\vert\vert^2_{HS} ds}$, $\overline{\vert\vert\ln \rho_s^D\vert\vert^2_{HS}} = \sqrt{\frac{1}{t} \int_0^{t} \vert\vert\ln \rho_s^D\vert\vert^2_{HS}ds }$, and $\overline{\vert\vert\ln \rho_s\vert\vert^2_{HS}} = \sqrt{\frac{1}{t} \int_0^{t} \vert\vert\ln \rho_s\vert\vert^2_{HS} ds } $. $\vert\vert \cdot \vert \vert_{HS}$ is the Hilbert-Schmidt norm, and $\mathcal{L}_s$ is the Liouvillian superoperator acting on $\rho$.%

\subsection{The model}
We consider an example of the decay of a two-state atom into a bosonic reservoir~\cite{Breuer_2012}. The  general form of the total Hamiltonian is given as 
\begin{equation}
    H = H_S \otimes \mathbb{I}_B + \mathbb{I}_S \otimes H_B + H_I, 
    \label{eq-total_ham}
\end{equation}
where $H_S$ is the system's Hamiltonian, $H_B$ is the Hamiltonian of the reservoir and interaction between the system and the reservoir is given by the Hamiltonian $H_I$. The form of the system's Hamiltonian is 
\begin{equation}
    H_S = \omega_0 \sigma_+\sigma_-, 
    \label{eq-system_ham}
\end{equation}
where $\sigma_+ = \ketbra{1}{0}$, and $\sigma_- = \ketbra{0}{1}$ are the atomic raising and lowering operators, respectively, with $\ket{0}(\ket{1})$ denoting the ground (excited) state. The environment is represented by a reservoir of harmonic oscillators given as 
\begin{equation}
    H_B = \sum_k \omega_k a_k^\dagger a_k,
\end{equation}
where $a_k$ and $a_k^\dagger$ are the bosonic creation and annihilation operators, satisfying the commutation relation $[a_k, a_{k'}^\dagger] = \delta_{k,k'}$. The form of the interaction Hamiltonian can be given as 
\begin{equation}
    H_{I} = \sum_{k} \left(g_k\sigma_+\otimes a_k + g^*_k\sigma_-\otimes a_k^{\dagger}\right),
    \label{eq-int_ham}
\end{equation}
where $g_k$ is the coupling constant. In this case, the total number of excitations in the system $N = \sigma_+\sigma_- + \sum_k a_k^\dagger a_k$ is a conserved quantity due to the rotating wave approximation. The dynamical map of the evolution of the reduced state of the system, with the bath initially in the vacuum state, was derived in~\cite{Breuer_1999} and is given by the quantum master equation
\begin{equation}
    \frac{d}{dt}\rho_{s}(t) = \frac{-i}{2} S(t) \left[\sigma_+\sigma_-, \rho_s\right] + \gamma(t)\left[\sigma_-\rho_s\sigma_+ - \frac{1}{2}\left\{\sigma_+\sigma_-, \rho_s\right\}\right], 
    \label{eq-NMAD_master_eq}
\end{equation}
where $\gamma(t) = -2\Re\left(\frac{\dot{G}(t)}{G(t)}\right)$, and $S(t) = -2\Im \left(\frac{\dot{G}(t)}{G(t)}\right)$. $\Re$ and $\Im$ represent real and imaginary parts of the quantity inside brackets, respectively. For a Lorentzian spectral density of the environment in resonance with the transition frequency of the qubit, the expression for the function $G(t)$ can be given as 
\begin{equation}
    G(t) = e^{-\lambda t/2}\left[\cosh\left(\frac{lt}{2}\right) + \frac{\lambda}{l}\sinh\left(\frac{lt}{2}\right)\right],
    \label{eq-func_G_t}
\end{equation}
where $l = \sqrt{\lambda^2 - 2\gamma_0\lambda}$. Here $\gamma_0$ describes the strength of the system-environment coupling, and $\lambda$ is the spectral width related to the environment. 
The quantities $S(t)$ and $\gamma(t)$ in Eq. (\ref{eq-NMAD_master_eq}) provide the time-dependent frequency shift and decay rates, respectively. Further, the quantity $-2\left(\frac{\dot{G}(t)}{G(t)}\right)$ can be written as 
\begin{equation}
    -2\frac{\dot G (t)}{G(t)} = 2\left(\frac{\gamma_0}{\sqrt{1 - \frac{2\gamma_0}{\lambda}\coth\left(\frac{1}{2}\lambda t\sqrt{1 - \frac{2\gamma_0}{\lambda}}\right) + 1}}\right).
\end{equation}
In the limit $\lambda<2\gamma_0$, the decay rate becomes negative for certain intervals giving rise to non-Markovian evolution, referred to as the non-Markovian amplitude damping (NMAD) evolution. In the limit $\lambda>2\gamma_0$, the dynamics become time-dependent Markovian. It can also be noted that for $\lambda \gg \gamma_0$, the decay rate $\gamma(t) = \gamma_0$, that is, it becomes time-independent, and the evolution corresponds to the standard AD channel. Further, using the Bloch vector representation of the single qubit density matrix, we can write the density matrix of the system at any time $t$ as
\begin{align}
\label{eq-Bloch_vec}
    \rho(t) = \frac{1}{2}\begin{pmatrix}
        1 + z(t) & x(t) - iy(t) \\
        x(t) + iy(t) & 1 - z(t)
    \end{pmatrix},
\end{align}
where $x(t) = {\rm Tr}[\sigma_x\rho(t)]$, $y(t) = {\rm Tr}[\sigma_y \rho(t)]$, and $z(t) = {\rm Tr}[\sigma_z\rho(t)]$. For the case considered above, the analytical expression for the dynamical map was derived in~\cite{Breuer_1999}, which is given by 
\begin{align}
\label{eq-dynamica_map}
    \rho_{00}(t) &= ( 1- |G(t)|^2)\rho_{11}(0) + \rho_{00}(0),\nonumber \\
    \rho_{01}(t) &= \rho_{01}(0)G^*(t),\nonumber \\
    \rho_{10}(t) &= \rho_{10}(0) G(t), \nonumber \\
    \rho_{11}(t) &= \rho_{11}(0) |G(t)|^2,
\end{align}
where $\rho_{ij}$ is the element of the density matrix $\rho(t)$, which we get as the solution of Eq. (\ref{eq-NMAD_master_eq}). A straightforward comparison between Eqs. (\ref{eq-dynamica_map}) and (\ref{eq-Bloch_vec}) gives us $x(t), y(t)$, and $z(t)$ in terms of the function $G(t)$. Thus, the Bloch vectors are given by 
\begin{align}
\label{eq-xt_yt_zt}
    x(t) &= 2\Re[\rho_{10}(0)G(t)], \nonumber \\
    y(t) &= -2\Im[\rho_{10}(0)G(t)], \nonumber \\
    z(t) &= 2\rho_{11}(0) |G(t)|^2 - 1.
\end{align}
One may note that the function $G(t)$ in Eq. (\ref{eq-func_G_t}) produces only real values due to its structure in both regimes when $l$ is real or imaginary. Therefore, the value of $y(t)$ in the above equation is zero for a real initial state $\rho(0)$. 

We can envisage the system as a quantum battery in the model studied here. The battery discharges when the ergotropy (discussed in the next subsection) of the system dissipates to the environment. We start with a state with finite ergotropy, and the ergotropy vanishes due to interaction with the environment during the evolution of the system. However, due to the system's P-indivisible non-Markovian evolution~\cite{nmad-channel}, the battery recharges again as the ergotropy revives. This is highlighted in the subsequent sections.

\begin{figure}[!htb]
    \centering
    \includegraphics[width = 1\columnwidth]{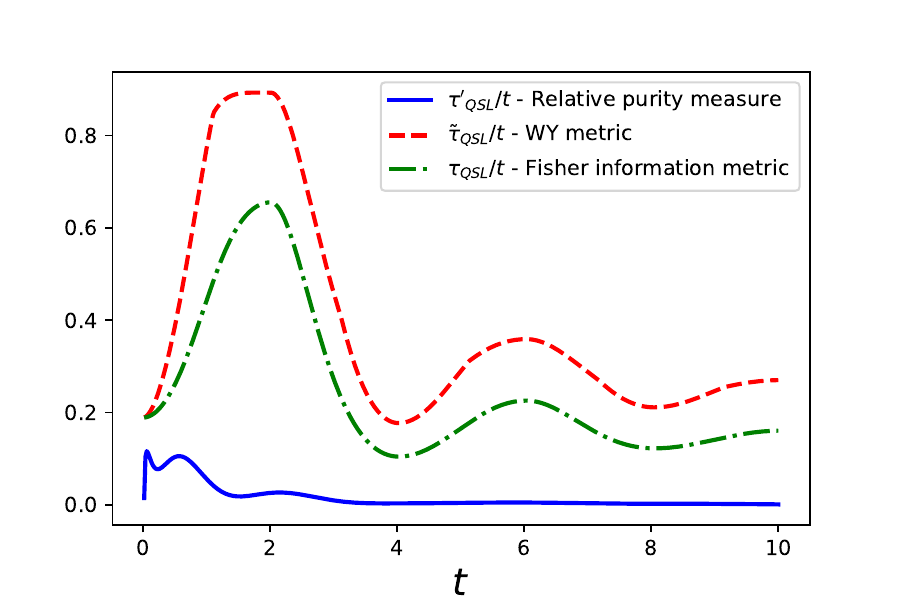}
    \caption{Comparison between QSL time obtained using Fisher information metric ($\tau_{QSL}$) WY information metric ($\tilde \tau_{QSL}$), and relative purity measure ($\tau'_{QSL}$) under the evolution of state through NMAD master equation. The parameters are: $\omega_0 = 1$, $\lambda = 0.5$, and $\gamma_0 = 10$.}
    \label{relative_purity_WY_Fisher_comparison_non_markov}
\end{figure}
In Fig. \ref{relative_purity_WY_Fisher_comparison_non_markov}, we compare the QSL time obtained using the Fisher information metric ($\tau_{QSL}$), Wigner-Yanase (WY) information metric ($\tilde \tau_{QSL}$), and the relative purity measure ($\tau'_{QSL}$). The evolution of the quantum state is through the NMAD master equation defined in Eq. (\ref{eq-NMAD_master_eq}). Here, we note that the QSL time using the Fisher information metric for this model was studied in~\cite{deffner2013qsl, Paulson2022}. The initial state, in this case, and subsequently is taken to be $\rho(0) = \ketbra{\psi_0}{\psi_0}$, where $\ket{\psi_0} = \frac{\sqrt{3}}{2}\ket{0} + \frac{1}{2}\ket{1}$, with $\ket{0}$ and $\ket{1}$ being the ground and excited state of the system, respectively. It can be observed that qualitatively the QSL time obtained using WY and Fisher information metrics are similar, with WY information giving a tighter bound on the QSL time. This is consistent with the comparison of both metrics in~\cite{adesso}.  

\subsection{Ergotropy, instantaneous and average power}
Here we define the ergotropy, followed by instantaneous and average powers. The categorization of ergotropy as a sum of incoherent and coherence ergotropies is also defined.
\subsubsection{Ergotropy}
Ergotropy refers to the maximum amount of work that can be extracted from a quantum system through a cyclic unitary transformation of the initial state. The problem of maximal extraction of work was discussed in~\cite{Allahverdyan_2004}. For a state governed with a time-dependent Hamiltonian $H_S + V(t)$, the time-dependent potential $V(t)$ corresponds to the transfer of work to external sources. The process is cyclic when the external source is connected at time $t=0$ and disconnected at time $t = t_0$, i.e., the time-dependent potential has the form $V(0) = V(t_0)=0$. One then looks out for the maximum work that can be extracted from the system for an arbitrary $V(t)$. To this effect, among all final states $\rho_{t_0}$ reached from the initial state $\rho_0$, we look for the state with the lowest final energy $E_f = {\rm Tr}[\rho_{t_0}H_S]$. The thermal equilibrium state $\rho^{eq}_{t_0} = \exp(-\beta H_S)/{\rm Tr}[\exp(-\beta H_S)]$ (where $\beta = T^{-1}$, with $T$ being the Temperature) under invariant von Neumann entropy is the standard answer to this problem. The largest amount of work that can be extracted in this case is $\mathcal{W}_{th} = E(\rho_0) - TS(\rho_0) + T\log(Z)$~\cite{Allahverdyan_2004}, where $S(\cdot)$ is the von Neumann entropy and $Z = {\rm Tr}[\exp(-\beta H_S)]$ is the partition function.

In the above, two arguments are made, one where we assume that the initial state can be brought to a thermal equilibrium state, and the other is that the entropy of the system is invariant during this evolution. It is allowed for the macroscopic systems where dissipative processes within the system may occur while the system evolves under the influence of $V(t)$. However, in the case of the finite systems, in general, only the action of potential $V(t)$ is not sufficient for a state to reach the thermal state in time $t_0$. Here, for the unitary evolution of the finite systems, not only the von Neumann entropy but the eigenvalues of $\rho$ are also conserved. Unlike thermodynamic systems, finite systems have the memory of their initial state, and relaxation mechanisms are not involved. Therefore, in these systems, the maximum amount of work extracted $\mathcal{W}$, called ergotropy, is generally smaller than $\mathcal{W}_{th}$. Below we outline the method to obtain ergotropy of the system.
We assume that we are given a quantum state $\rho_0$ with its internal Hamiltonian $H_S$ such that they have the following spectral decomposition:
\begin{equation}
    \rho_0 = \sum_{i} r_i\ket{r_i}\bra{r_i},
\end{equation}
 and 
 \begin{equation}
     H_S = \sum_{i} \epsilon_i\ket{\epsilon_i}\bra{\epsilon_i},
 \end{equation}
where ordering of the eigenvalues for $\rho_0$ and $H_S$ is in the decreasing, $r_1\ge r_2\ge....,$ and increasing, $\epsilon_1\le \epsilon_2\le...$, order, respectively. Since unitary dynamics is considered, any decrease in the internal energy of the system at hand, with respect to its self-Hamiltonian $H_S$, will be extracted as work. Thus, in order to find the ergotropy, one aims to minimize the internal energy of the final state
\begin{equation}
    \mathcal{W}(\rho_0) = {\rm Tr}(\rho_0 H_S) - \min\{{\rm Tr} (U\rho U^{\dagger} H_S)\},
    \label{eq-ergotropy_1}
\end{equation}
where the minimization is performed over all possible unitaries.

It can be shown that the final state $\rho_{t_0} = \rho_f = U \rho U^{\dagger}$ that achieves this minimum should commute with $H_S$ and have the same eigenvalues as $\rho_0$, that is, $\rho_f = \sum_j r_j \ket{\epsilon_j}\bra{\epsilon_j}$. This is known as the passive state because no work can be extracted from this state. The intuition for the order of eigenvalues comes from the interpretation that the highest occupation fraction $r_1$ of $\rho_0$ should occupy the lowest level. A unitary operator which performs such a transformation is $U = \sum_j\ket{\epsilon_j}\bra{r_j}$. 
Now we can rewrite Eq. (\ref{eq-ergotropy_1}) as 
\begin{equation}
    \mathcal{W}(\rho_0) = \sum_{j, i} r_j\epsilon_i (|\bra {r_j} \epsilon_i\rangle|^2 - \delta_{ij}).
    \label{eq-ergotropy_2}
\end{equation}
The ergotropy $\mathcal{W}$ depends only on the initial state and Hamiltonian of the system. The upper bound on ergotropy is given by $\mathcal{W}_{th}\ge \mathcal{W}\ge0$, where $\mathcal{W}_{th}$ is equal to $\mathcal{W}$ when $r_j = -\beta\epsilon_j - \log(Z)$. Furthermore, ergotropy has already been studied for open quantum systems in~\cite{cakmak_ergo, Kamin_2020, Touil_2022}. In the previous section, we discussed a simple open quantum system modeling a non-Markovian amplitude damping channel. To calculate the ergotropy of the system, we make use of the solution $\rho(t)$ obtained after solving Eq. (\ref{eq-NMAD_master_eq}) for any time $t$ and supply it as the initial state in the Eq. (\ref{eq-ergotropy_1}) to find the maximum work that can be extracted from this state after a cyclic unitary transformation. Physically, the work is extracted from the system after it is detached from the bath and is now subjected to the transformation stated above.

For the state $\rho(t)$ given in Eq. (\ref{eq-Bloch_vec}), one can analytically calculate the ergotropy of the system with inherent system Hamiltonian $H_S$ defined by Eq. (\ref{eq-system_ham}). To this end, the eigenvalues of the state $\rho(t)$ are $\frac{1}{2}(1 \pm |a(t)|)$, where $|a(t)| = \sqrt{x(t)^2 + z(t)^2}$. Note that we have dropped $y(t)$ from the calculations because it is zero for the case considered here, as given in Eq. (\ref{eq-xt_yt_zt}). The spectral decomposition of the $H_S = \omega_0\ket{1}\bra{1} + 0\ketbra{0}{0}$ can now be used to find the passive state $\rho_f(t) = r_1\ketbra{0}{0} + r_2\ketbra{1}{1}$, where $2r_1 = 1+|a(t)| \ge 1-|a(t)| = 2r_2$. The ergotropy of the system now boils down to 
\begin{align}
\label{eq-analytical_ergo1}
 \mathcal{W}(\rho(t)) &= {\rm Tr}(\rho(t) H_S) - {\rm Tr}(\rho_f(t) H_S), \nonumber \\
 &= \omega_0\left(\frac{1+z(t)}{2}\right) - \omega_0\left(\frac{1-|a(t)|}{2}\right) = \frac{\omega_0}{2}\left[|a(t)| + z(t)\right], \nonumber \\ 
 &= \frac{\omega_0}{2}\left(\sqrt{x(t)^2 + z(t)^2} + z(t)\right).
\end{align}
In the above equation, $z(t)$ denotes the population terms, and $x(t)$ denotes the coherence terms of the state at any time $t$ (Eq. (\ref{eq-Bloch_vec})). The value of the ergotropy is non-zero whenever there is coherence term $x(t)$ present in the state. It becomes zero only when $x(t)$ is zero and $z(t)$ is less than or equal to zero. Using the Eq. (\ref{eq-xt_yt_zt}) and the initial state $\rho(0) = \ketbra{\psi_0}{\psi_0}$, where $\ket{\psi_0} = \frac{\sqrt{3}}{2}\ket{0} + \frac{1}{2}\ket{1}$, we can write the ergotropy in terms of the function $G(t)$ as 
\begin{equation}
\label{eq-analytical_ergo2}
    \mathcal{W}(\rho(t)) = \frac{\omega_0}{4}\left(-2 + G(t)^2 +\sqrt{4 - G(t)^2 + G(t)^4}\right).
\end{equation}%

It was recently recognized~\cite{Francica_coherent_ergo, Deffner_coherent_ergo} that quantum ergotropy can be separated into two different (coherent and incoherent) contributions 
\begin{equation}
    \mathcal{W} = \mathcal{W}_i + \mathcal{W}_c.
    \label{eq-coherent-incoherent-ergos}
\end{equation}
The incoherent ergotropy $\mathcal{W}_i$ denotes the maximal work that can be extracted from $\rho$ without changing its coherence, which is defined as
\begin{equation}
    \mathcal{W}_i(\rho) = {\rm Tr} \left\{\left(\rho - \sigma \right)H_S\right\},
    \label{eq-incoherent-ergo}
\end{equation}
where $\sigma$ is the coherence invariant state of $\rho$, which has property 
\begin{equation}
    {\rm Tr} \left\{\sigma H_S\right\} = \min_{\mathcal{U} \in \mathcal{U}^{(i)}} {\rm Tr} \left\{\mathcal{U}\rho\mathcal{U}^\dagger H_S\right\},
    \label{eq-coherent_invariant_state}
\end{equation}
where $\mathcal{U}^{(i)}$ is the set of unitary operations without changing the coherence of $\rho$.
Alternatively, the incoherent ergotropy can be calculated using the state $\rho$ after erasing all its coherence terms by applying a dephasing map and then using the same method used in calculating the full ergotropy for this dephased state. That is, consider the state $\rho^D = \sum_i\braket{i|\rho|i}\ketbra{i}{i}$; the incoherent ergotropy $\mathcal{W}_i(\rho)$ for state $\rho$ is equivalent to the ergotropy $\mathcal{W}$ of the state $\rho^D$. The passive state $\rho_f^D$, in this case, can be found in a similar way as done previously for the calculation of ergotropy. Therefore, 
\begin{equation}
    \mathcal{W}_i(\rho) = \mathcal{W}(\rho^D) = {\rm Tr}\left[\left(\rho^D - \rho^D_f\right)H_S\right].
\end{equation}
The dephased state $\rho^D(t)$ for the state $\rho(t)$ in Eq. (\ref{eq-Bloch_vec}) is given by $\frac{1}{2}\left[(1 + z(t))\ketbra{1}{1} + (1 - z(t))\ketbra{0}{0}\right]$, and the corresponding passive state $\rho_f^D(t)$ depends on the sign of $z(t)$. Considering the case $z(t)<0$, we find that $1 + z(t)$ is lesser than $1 - z(t)$. To this end, the passive state $\rho^f_D(t)$ is found to be the same as $\rho^D(t)$; therefore, the incoherent ergotropy $\mathcal{W}_i(\rho(t))$ becomes zero. However, if $z(t)\ge0$, then $1+z(t)$ is greater than $1-z(t)$ and $\rho_f^D(t) = \frac{1}{2}\left[(1 - z(t))\ketbra{1}{1} + (1 + z(t))\ketbra{0}{0}\right]$. In this case, for the system Hamiltonian $H_S = \omega_0\ketbra{1}{1}$, the incoherent ergotropy is given by 
\begin{align}
    \mathcal{W}_i(\rho(t))= \begin{cases}
    \omega_0 z(t). & \text{for } z(t) \ge 0\\
    0. & \text{for } z(t) < 0
    \end{cases}
\end{align}
Upon comparing the above equation with Eq. (\ref{eq-analytical_ergo1}), we observe that for positive $z(t)$ and zero coherence term $x(t)/2$, the incoherent ergotropy $\mathcal{W}_i$ becomes equal to the ergotropy $\mathcal{W}$. Further, the incoherent ergotropy is always positive here and gets contribution only from the population terms of $\rho(t)$. The incoherent ergotropy in terms of the function $G(t)$ and the initial state $\rho(0)$ defined above Eq. (\ref{eq-analytical_ergo2}), is given by  
\begin{align} 
    \mathcal{W}_i(\rho(t))= \begin{cases}
    \omega_0 \frac{G(t)^2 -2}{2}. & \text{for } G(t)^2  \ge 2\\
    0. & \text{for } G(t)^2 < 2
    \end{cases}
\end{align}

Moreover, it is straightforward now to get the expression for the coherent ergotropy $\mathcal{W}_c = \mathcal{W} - \mathcal{W}_i$ for the state $\rho(t)$ given in Eq. (\ref{eq-Bloch_vec}) and system Hamiltonian $H_S$ in Eq. (\ref{eq-system_ham}). This is given by 
\begin{align}
\label{eq-analytical_coherent_ergo}
    \mathcal{W}_c(\rho(t)) = \begin{cases}
    \frac{\omega_0}{2}\left(\sqrt{x(t)^2 + z(t)^2} - z(t)\right). & \text{for } z(t) \ge 0\\
    \frac{\omega_0}{2}\left(\sqrt{x(t)^2 + z(t)^2} + z(t)\right). & \text{for } z(t) < 0
    \end{cases}
\end{align}%
The coherent ergotropy $\mathcal{W}_c$ is the work that is exclusively stored in the coherence. his can be verified from the above equation. We note here that whenever the coherence term $x(t)/2$ of the state $\rho(t)$ is zero, the coherent ergotropy of the system also becomes zero. Also, the coherent ergotropy becomes the full ergotropy for the negative values of $z(t)$. The expression of coherent ergotropy $\mathcal{W}_c$ in terms of the function $G(t)$ and for initial state $\rho(0)$ can be obtained in a similar manner as done for ergotropy and incoherent ergotropy using Eq. (\ref{eq-xt_yt_zt}). Furthermore, we outline the relationship between the coherent ergotropy and the relative entropy of coherence $C(\rho)$ defined in Eq. (\ref{eq-coherence_1}). The expression was derived in~\cite{Francica_coherent_ergo} and can be given via quantum relative entropy as  
%This can be quantified by the relative entropy of coherence defined in Eq. (\ref{eq-coherence_1}). The expression for the coherent ergotropy was derived in~\cite{Francica_coherent_ergo} and can be written in terms of classical relative entropy as 
\begin{equation}
    \beta \mathcal{W}_c (\rho) = C(\rho) + D\left(\rho^D_f \vert \vert \rho^{eq}\right) - D\left(\rho_f \vert\vert \rho^{eq}\right),
    \label{eq-coherent_ergo}
\end{equation}
where $D(\sigma ||\rho) = {\rm Tr}\left[\sigma\left(\log\sigma - \log \rho\right)\right]$ is the quantum relative entropy. $\rho^D_f$ is the passive state of the dephased state $\rho^D$ and $\rho_f$ is the passive state of $\rho$. 
The state $\rho^{eq}$ is the Gibbs state 
\begin{equation}
    \rho^{eq} = \frac{\exp(-\beta H_S)}{Z}\,\,\,\, \text{with} \,\,\,\, Z = {\rm Tr} \left\{\exp(-\beta H_S)\right\},
\end{equation}
and $\beta = 1/k_B T$, with $T$ being the temperature. 
Despite the fact that in Eq. (\ref{eq-coherent_ergo}), temperature $T$ is present explicitly, yet upon substituting the values of all the terms in the right-hand side using Eq. (\ref{eq-Bloch_vec}) and dividing with $\beta$, we get the same expression as in Eq. (\ref{eq-analytical_coherent_ergo}), i.e., free of $T$.

We next establish a relation between the coherent ergotropy and another quantifier of the coherence of a quantum state given by $l_1$ norm of coherence~\cite{coherence_l1_norm}, defined as 
\begin{equation}
   \mathcal{C}_{l_1}(\rho) = \sum_{i, j, i\ne j} \vert \rho_{i, j}\vert,
   \label{eq-coherence_l_1_norm}
\end{equation}
For the state of the form of Eq. (\ref{eq-Bloch_vec}) with $y(t)$ being zero, the value of $\mathcal{C}_{l_1}(\rho(t))$ is given by
\begin{align}
\label{analytical_l1_norem}
    \mathcal{C}_{l_1}(\rho(t)) = |x(t)|.
\end{align}
Using the above equation and Eq. (\ref{eq-analytical_coherent_ergo}), for the case studied here, we can write coherent ergotropy $\mathcal{W}_c$ in terms of $\mathcal{C}_{l_1}(\rho(t))$ as
\begin{align}
    \mathcal{W}_c(\rho(t)) = \begin{cases}
    \frac{\omega_0}{2}\left(\sqrt{\mathcal{C}_{l_1}(\rho(t))^2 + z(t)^2} - z(t)\right). & \text{for } z(t) \ge 0\\
    \frac{\omega_0}{2}\left(\sqrt{\mathcal{C}_{l_1}(\rho(t))^2 + z(t)^2} + z(t)\right). & \text{for } z(t) < 0
    \end{cases}
\end{align}
We note that in case the value of $z(t)$ becomes zero, the coherent ergotropy $\mathcal{W}_c$ gets directly proportional to the value of $\mathcal{C}_{l_1}(\rho(t))$.

\subsubsection{Average and instantaneous power}
The instantaneous charging power is defined by available work in the battery as
\begin{equation}
    \mathcal{P}(t) = \lim_{\Delta t\rightarrow 0} \frac{\mathcal{W}(t + \Delta t) - \mathcal{W}(t)}{\Delta t} = \frac{d\mathcal{W}}{dt}.
\end{equation}
Using Eq. (\ref{eq-analytical_ergo1}), we can write the instantaneous power $\mathcal{P}(t)$ as 
\begin{align}
    \mathcal{P}(t) = \frac{\omega_0}{2}\left(\frac{x(t)\dot x(t) + z(t) \dot z(t)}{\sqrt{x(t)^2 + z(t)^2}} + \dot z(t)\right).
\end{align}%
It is also possible to define the average power-to-energy transfer given by 
\begin{equation}
    \mathcal{P}_{av} = \frac{\mathcal{W} (t) - \mathcal{W} (t_0)}{t - t_0},
\end{equation}
where $t-t_0$ refers to the charging time of the battery.

We will now see the connection between various forms of the quantum speed limit, particularly QSL time for coherence and using relative purity measure, with the ergotropy, with the dynamics being generated by the NMAD model.
\section{Connection of the quantum speed limits with the coherent ergotropy}\label{sec-coherent-connect}
This section first discusses the connection between the QSL for coherence and coherent ergotropy. Further, we connect the QSL time obtained using the relative purity measure and the coherent ergotropy. 

\subsection{Connection between the QSL for coherence and the coherent ergotropy}
\begin{figure}[!htb]
    \centering
    \includegraphics[width = 1\columnwidth]{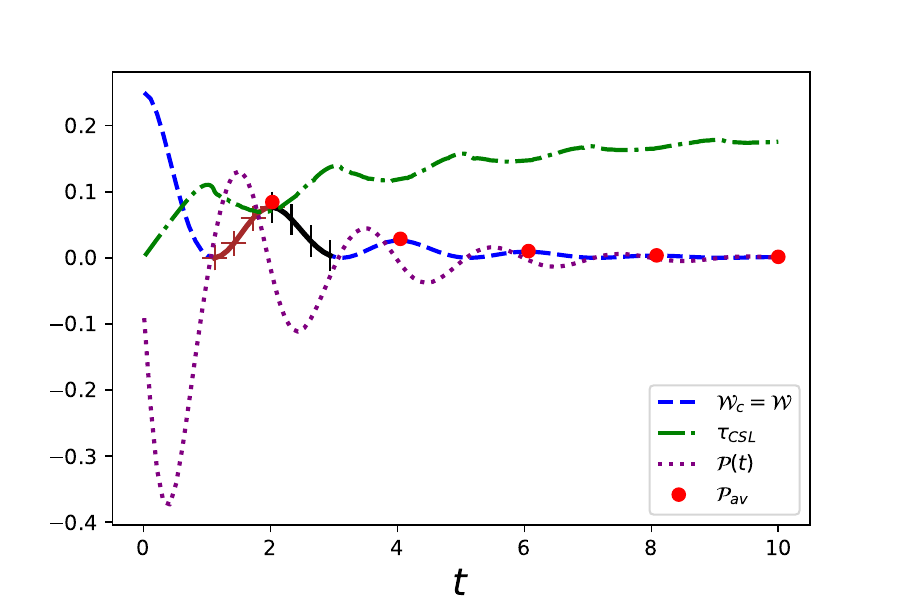}
    \caption{Variation of the QSL time for coherence $\tau_{CSL}$, ergotropy $\mathcal{W}$, coherent ergotropy $\mathcal{W}_C$, instantaneous and average powers, $\mathcal{P}(t)$ and $\mathcal{P}_{av}$, respectively, with time. The state is evolved through the NMAD master equation. The parameters are chosen to be $\omega_0 = 1$, $\lambda = 0.5$, and $\gamma_0 = 10$. }
    \label{fig-qsl_coherent_ergotropy_power}
\end{figure}

We can rewrite Eq. (\ref{eq-coherent_ergo}) for the relative entropy of coherence in terms of coherent ergotropy as 
\begin{equation}
    C(\rho) = \beta \mathcal{W}_c (\rho)  - D\left(\rho^D_f \vert \vert \rho^{eq}\right) + D\left(\rho_f \vert\vert \rho^{eq}\right).
    \label{eq-coherent_ergo2}
\end{equation}
Plugging the above equation in Eq. (\ref{eq-qsl_coherence}) for the state $\rho(t)$ at time $t$ and at time $t = 0$, we get the following relation between the QSL time for coherence $\tau_{CSL}$ and the coherent ergotropy $\mathcal{W}_c$ 
\begin{equation}
    \tau_{CSL} = \frac{\vert \beta \mathcal{W}_c (\rho(t)) - D\left(\rho^D_f(t) \vert \vert \rho^{eq}\right) + D\left(\rho_f(t) \vert\vert \rho^{eq}\right) - C(\rho(0))\vert}{\Lambda^{\text{rms}, D}_{t} \overline{\vert\vert\ln \rho_s^D\vert\vert^2_{HS}} + \Lambda^{\text{rms}}_{t}\overline{\vert\vert\ln \rho_s\vert\vert^2_{HS}}}, 
    \label{eq-qsl_ergo_relation}
\end{equation}
where $C(\rho(0)) = \beta \mathcal{W}_c (\rho(0)) - D\left(\rho^D_f(0) \vert \vert \rho^{eq}\right) + D\left(\rho_f(0) \vert\vert \rho^{eq}\right)$. Here, and subsequently, in the next sections, we take the initial state to be $\rho(0) = \ketbra{\psi_0}{\psi_0}$, where $\ket{\psi_0} = \frac{\sqrt{3}}{2}\ket{0} + \frac{1}{2}\ket{1}$. The value of $C(\rho(0))$ for this initial state is around $0.56$. Using the Bloch vector representation of the state $\rho(t)$ (Eq. (\ref{eq-Bloch_vec})), the relative entropy of coherence $C(\rho(t))$ is given by 
\begin{align}
\label{eq-analytical_relative_coherence}
    C(\rho(t)) &= -\left(\frac{1+z(t)}2\right)\log\left(\frac{1+z(t)}{2}\right)- \left(\frac{1-z(t)}2\right)\log\left(\frac{1-z(t)}{2}\right) \nonumber \\
    &+\left(\frac{1+|a(t)|}2\right)\log\left(\frac{1+|a(t)|}{2}\right) + \left(\frac{1-|a(t)|}2\right)\log\left(\frac{1-|a(t)|}{2}\right),
\end{align}
where $|a(t)| = \sqrt{x(t)^2 + z(t)^2}$, as $y(t)$ is zero for the case considered here. We note here that for the given initial state $\rho(0)$ above, the value of $z(t)$ remains negative for the whole dynamics. Interestingly, in this case, the value of ergotropy $\mathcal{W}$ and coherent ergotropy $\mathcal{W}_c$ is exactly the same. Therefore, the relationship between the coherent ergotropy $\mathcal{W}_c$ and $\tau_{CSL}$ corresponds to a relationship between ergotropy $\mathcal{W}$ and $\tau_{CSL}$ too. Further, the relative entropy of coherence $C(\rho(t))$ vanishes whenever $x(t)$ is zero, and at those points in time, the ergotropy $\mathcal{W}$ and coherence ergotropy $\mathcal{W}_c$ are also zero. This is depicted in Fig. \ref{fig-qsl_coherent_ergotropy_power}. At these points in time, the $\tau_{CSL}$ achieves a local maximum, which is intuitive because the numerator of $\tau_{CSL}$ is locally maximum.%
In Fig. \ref{fig-qsl_coherent_ergotropy_power}, we can see the variation of the QSL time for coherence ($\tau_{CSL}$) and the characterizers of quantum thermodynamics, particularly ergotropy ($\mathcal{W}$), coherent ergotropy ($\mathcal{W}_C$), instantaneous ($\mathcal{P}(t)$) and average ($\mathcal{P}_{av}$) powers. The region in the ergotropy curve denoted by the brown $``+"$ symbols indicates the battery's charging process, while the black $``|"$ symbols depict the discharging process in the given cycle from time $t =1$ to 3. During the charging process, the instantaneous power is positive and negative while the battery is discharging. Further, in this case, we observe that the maxima and minima of the $\tau_{CSL}$ are in contrast with the erogtropy and coherent ergotropy's maxima and minima. This shows that the speed of evolution of the coherence of the state during the discharging process is getting slower. This speed hits its minimum when the ergotropy is zero. However, during the charging process, when ergotropy again becomes non-zero, the speed of evolution of the coherence increases and reaches its maximum when the battery is fully charged. We also point out here that one can take a different initial state $\rho(0)$ and can get some contribution to ergotropy from the incoherent ergotropy too. However, the dynamics of the system characterized by the NMAD master equation quickly drives the system's state in such a way that the value of $z(t)$ becomes negative. In this case, at longer times, the coherent ergotropy eventually becomes equal to the ergotropy of the system.

\subsection{QSL time for relative purity measure and coherent ergotropy}
Here, we start with the  Eq. (\ref{eq-qsl_rel_purity}) using the initial state $\rho(0)$ used in the previous section and the Bloch vector form of the system's state $\rho(t)$ given in Eq. (\ref{eq-Bloch_vec}) at any time $t$. To this end, the value of ${\rm Tr}[\rho(0)^2]$ is one as we have considered an initial pure state. The value of the relative purity $f(t) = {\rm Tr}[\rho(t)\rho(0)]/{\rm Tr}[\rho(0)^2]$ is now given by
\begin{align}
    f(t) = \frac{1}{4}\left(2 - z(t) + \sqrt{3}x(t)\right).
\end{align}
Further, the value of the term $\sqrt{{\rm Tr}\left[\left(\mathcal{L}^\dagger \rho(0)\right)^2\right]}$ present in the denominator of Eq. (\ref{eq-qsl_rel_purity}) becomes $\sqrt{\frac{11}{8}}\left|\frac{G(s)}{\dot G(s)}\right|$, where the function $G(s)$ is given in Eq. (\ref{eq-func_G_t}). Now the QSL time obtained using the relative purity measure $\tau'_{QSL}$ boils down to 
\begin{align}
    \tau'_{QSL} = \frac{8\sqrt{\frac{2}{11}}\left[\cos^{-1}\left(\frac{1}{4}\left(2 - z(t) + \sqrt{3}x(t)\right)\right)\right]^2}{\frac{\pi^2}{t}\int_0^t\left|\frac{G(s)}{\dot G(s)}\right|dt}.
\end{align}
Here, we again mention that the value of $z(t)$ for the given initial state $\rho(0)$ is always negative. Therefore, the incoherent ergotropy is zero, and the coherent ergotropy is equal to the ergotropy of the system. Further, we observe that in contrast to $\tau_{CSL}$'s variation of the numerator, in this case, when the coherence term $x(t)$ and the ergotropy and coherent ergotropy goes to zero, the numerator obtains a minimum value.% 
\begin{figure}[!htb]
    \centering
    \includegraphics[width = 1\columnwidth]{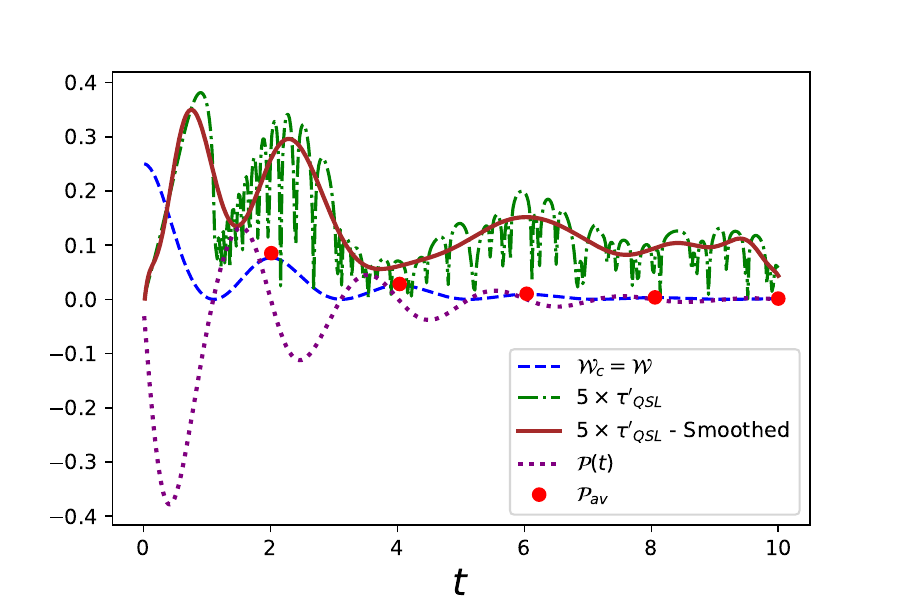}
    \caption{Variation of QSL time with relative purity measure ($\tau'_{QSL}$), ergotropy ($\mathcal{W}$), coherent ergotropy ($\mathcal{W}_c$), instantaneous ($\mathcal{P}(t)$) and average ($\mathcal{P}_{av}$) powers, with evolution time $t$. The evolution is through the NMAD master equation. The parameters are taken to be $\omega_0 = 1$, $\lambda = 0.5$, and $\gamma_0 = 10$. Note that we have scaled the value of $\tau'_{QSL}$ by multiplying a constant factor of 5 for a better comparison of its variation with other quantities in the figure.}
    \label{fig-qsl_relative_purity_ergotropy_power}
\end{figure}

Figure \ref{fig-qsl_relative_purity_ergotropy_power} depicts the connection between the QSL time using relative purity measure ($\tau'_{QSL}$) with ergotropy ($\mathcal{W}$), coherent ergotropy ($\mathcal{W}_c$), instantaneous ($\mathcal{P}(t)$), average ($\mathcal{P}_{av}$) powers. It can be observed that, in this case, the peaks and valleys of $\tau'_{QSL}$ are in contrast with the peaks and valleys of instantaneous power. This indicates that changes in the $\tau'_{QSL}$ pick up the maximal rate of charging or discharging of the battery, with a maximum in $\tau'_{QSL}$ corresponding to the maximum discharging rate of the battery and a minimum in $\tau'_{QSL}$ corresponds to the maximum charging rate, in a given charging and discharging cycle.  

\section{Quantum speed limit and Ergotropy in non-Markovian evolution}\label{sec-qsl-ergo-connect}
\begin{figure}[!htb]
    \centering
    \includegraphics[width = 1\columnwidth]{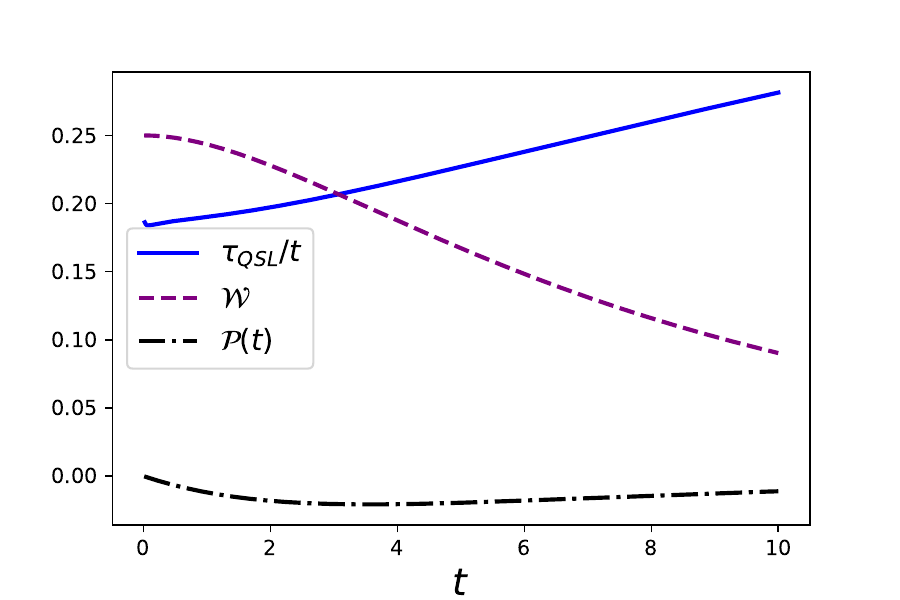}
    \caption{Variation of QSL time using Fisher information metric ($\tau_{QSL}$), ergotropy ($\mathcal{W}$), instantaneous ($\mathcal{P}(t)$) and average ($\mathcal{P}_{av}$) powers, with time. The evolution of the state is through the Markovian AD channel. The parameters are taken to be $\omega_0 =1 $, $\lambda = 0.5$, $\gamma_0 = 0.1$.}
    \label{fig-Fisher_qsl_ergo_markov}
\end{figure}

Now we discuss the impact of the non-Markovian evolution on the charging and discharging process of the quantum battery together with its impact on the QSL time, particularly those obtained using geometric measures. 
In Fig. \ref{fig-Fisher_qsl_ergo_markov}, the evolution of the system under Markovian evolution is depicted. 
Here we observe that the system (qubit) keeps on discharging, but due to the Markovian evolution, there is no revival in the ergotropy or charging of the battery. This brings out the fact that in the present qubit-bath setup, the non-Markovian nature of the evolution plays a crucial role in highlighting the role of the system as a battery. Further, the QSL time obtained using the Fisher information metric ($\tau_{QSL}$) is also monotonically increasing; that is, the speed of evolution of the system keeps slowing down with time.  
\begin{figure}[!htb]
    \centering
    \includegraphics[width = 1\columnwidth]{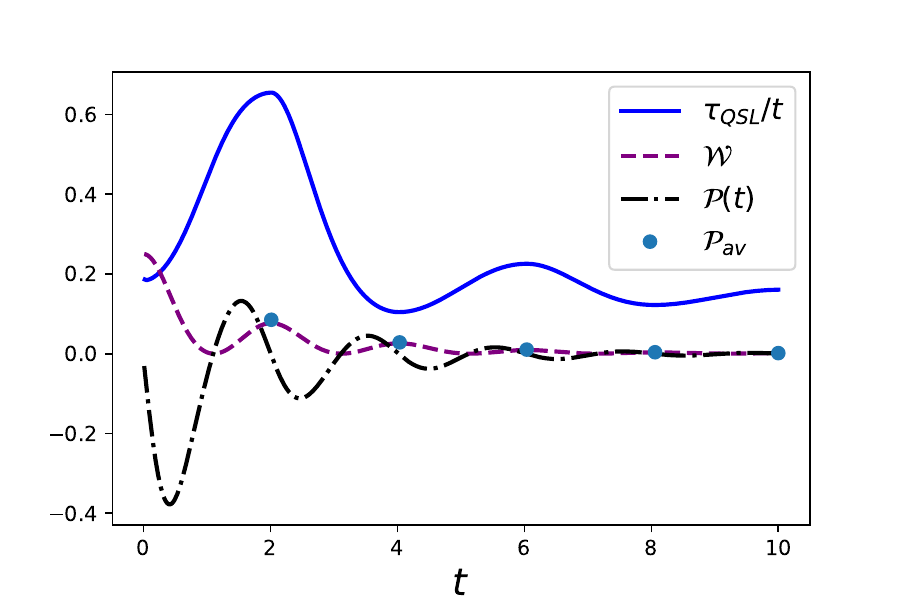}
    \caption{Variation of QSL time using Fisher information metric ($\tau_{QSL}$), ergotropy ($\mathcal{W}$), and instantaneous ($\mathcal{P}(t)$) and average ($\mathcal{P}_{av}$) powers, with time. The evolution of the state is through the NMAD channel. The parameters are taken to be $\omega_0 =1 $, $\lambda = 0.5$, $\gamma_0 = 10$.}
    \label{fig-Fisher_qsl_ergo_non_markov}
\end{figure}

\begin{figure}[!htb]
    \centering
    \includegraphics[width = 1\columnwidth]{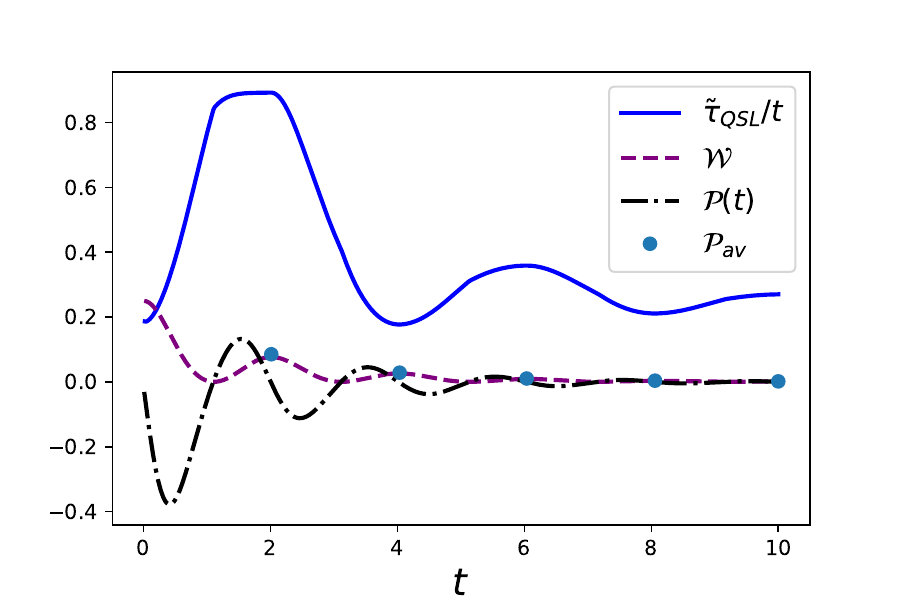}
    \caption{Variation of QSL time using Wigner-Yanase information metric ($\tilde \tau_{QSL}$), ergotropy ($\mathcal{W}$), and instantaneous ($\mathcal{P}(t)$) and average ($\mathcal{P}_{av}$) powers, with time. The evolution of the state is through the NMAD channel. The parameters are taken to be $\omega_0 =1 $, $\lambda = 0.5$, $\gamma_0 = 10$.}
    \label{fig-WY_qsl_ergo_non_markov}
\end{figure}

Figures \ref{fig-Fisher_qsl_ergo_non_markov} and \ref{fig-WY_qsl_ergo_non_markov} show the variation of the QSL time obtained using geometric measures (using Fisher and Wigner-Yanase information metrics), ergotropy, instantaneous and average powers, with time. 
Qualitatively the QSL time obtained using Fisher ($\tau_{QSL}$) and WY ($\tilde \tau_{QSL}$) information metrics show a similar pattern. In both cases, we observe that either of the peaks or valleys of the $\tau_{QSL}$ and $\tilde \tau_{QSL}$ occurs exactly at the points when a cycle of complete discharging and recharging to a maximum value complete. This coincides with the points where the average power is calculated. The revivals in the ergotropy (or recharging of the battery) are completely due to the non-Markovian nature of the system in both figures. This also brings out a difference between the QSL time obtained using inherent dynamics of the system, particularly QSL time for coherence ($\tau_{CSL}$ in Fig. \ref{fig-qsl_coherent_ergotropy_power})  and using relative purity measure ($\tau'_{QSL}$ in Fig. \ref{fig-qsl_relative_purity_ergotropy_power}), and the QSL time obtained using geometric measures in Figs. \ref{fig-Fisher_qsl_ergo_non_markov} and \ref{fig-WY_qsl_ergo_non_markov}. The peaks and valleys of the $\tau_{CSL}$ and $\tau'_{QSL}$ specify the discharging and charging processes, being exact for $\tau_{CSL}$. However, the peaks and valleys of $\tau_{QSL}$ and $\tilde \tau_{QSL}$ specify the completion of a discharging-charging cycle. This thus brings out a difference between the geometric and physical properties-based measures of QSL time from the perspective of discharging-charging processes. It can also be seen that the speed of evolution first decreases and then increases, oscillating around a fixed point later. This again benchmarks the non-Markovian nature of the system. 

\section{Conclusions}\label{sec-conclusions}
Here we have studied the impact of non-Markovian evolution on the quantities characterizing quantum thermodynamics, particularly ergotropy and its components, and instantaneous and average powers. We explored the connection of these quantities with the quantum speed limit (QSL) time of evolution. The characterization of the QSL time was done using both geometric and physical properties-based measures. Fisher and Wigner-Yanase information metrics were used for the geometric-based measures ($\tau_{QSL}$ and $\tilde \tau_{QSL}$, respectively), and relative purity and relative entropy of coherence were used for the physical properties-based measure ($\tau_{CSL}$ and $\tau'_{QSL}$, respectively) of QSL time. We considered the evolution of a single qubit system interacting with the bosonic bath through the non-Markovian amplitude damping (NMAD) master equation. Having an initial finite ergotropy, we proposed that the system could be envisaged as a quantum battery discharging into the bosonic environment. 

The coherent and incoherent components of erogtropy were discussed. It was observed that due to the nature of the non-Markovian amplitude damping evolution of the state, the contribution to the ergotropy in the form of incoherent ergotropy quickly vanishes, and the coherent ergotropy becomes equivalent to the ergotropy of the system. Further, direct connections between the coherent ergotropy and QSL time using relative purity and relative entropy of coherence measures were explored. It was observed that QSL time obtained using physical properties-based measures identified the discharging and charging process of the quantum battery. This was different from the geometric-based measures that brought out the connection between the completion of the discharging-charging cycle and the change in the speed of evolution of the system. The revivals in the ergotropy of the system (or recharging of the battery) only occurred in the non-Markovian limit. It was observed that in the Markovian limit, the battery only discharged, and the speed of evolution was monotonically decreasing (QSL time was monotonically increasing). Therefore, the non-Markovian evolution is crucial when modeling the system as a quantum battery. 

\section*{Acknowledgements}
SB acknowledges support from the Interdisciplinary Cyber-Physical Systems (ICPS) programme of the Department of Science and Technology (DST), India, Grant No.: DST/ICPS/QuST/Theme-1/2019/13. SB also acknowledges support from the Interdisciplinary Research Platform (IDRP) on Quantum Information and Computation (QIC) at IIT Jodhpur.

\bibliographystyle{apsrev4-1}
\bibliography{reference}

\end{document}